\documentclass[conference]{IEEEtran}
\IEEEoverridecommandlockouts
\usepackage{cite}
\usepackage{amsmath,amssymb,amsfonts}
\usepackage{algorithmic}
\usepackage{graphicx}
\usepackage{textcomp}
\usepackage{comment}
\usepackage{xcolor}
\def\BibTeX{{\rm B\kern-.05em{\sc i\kern-.025em b}\kern-.08em
    T\kern-.1667em\lower.7ex\hbox{E}\kern-.125emX}}
\begin{document}

\title{Dynamic EM Ray Tracing for Large Urban Scenes with Multiple Receivers\\

}

\author{\IEEEauthorblockN{Ruichen Wang}
\IEEEauthorblockA{\textit{ECE Department} \\
\textit{University of Maryland, College Park}\\
Maryland, USA \\
rwang92@umd.edu}
\and
\IEEEauthorblockN{Dinesh Manocha}
\IEEEauthorblockA{\textit{CS and ECE Department} \\
\textit{University of Maryland, College Park}\\
Maryland, USA \\
dmanocha@umd.edu}
}

\maketitle

\begin{abstract}
Radio applications  are increasingly being used in urban environments for cellular radio systems and safety applications that use vehicle-vehicle, and vehicle-to-infrastructure. We present a novel ray tracing-based radio propagation algorithm that can handle large urban scenes with hundreds or thousands of  dynamic objects and receivers. Our approach is based on the use of coherence-based  techniques that exploit spatial and temporal coherence for efficient wireless propagation and radio network planning. 
Our formulation also utilizes channel coherence which is used to determine the effectiveness of the propagation model within a certain time in dynamically generated paths; and spatial consistency which is used to estimate the similarity and accuracy of changes in a dynamic environment with varying propagation models and blocking obstacles. We highlight the performance of our simulator in large urban traffic scenes with an area of $2*2 km^2$ and more than 10,000 users and devices. We evaluate the accuracy by comparing the results with discrete model simulations performed using WinProp. In practice, our approach scales linearly with the area of the urban environment and the number of dynamic obstacles or receivers. 
\end{abstract}


\section{Introduction}
Ray tracing propagation models have been developed over the last four decades  for visual, aural, and electromagnetic (EM) simulations. In terms of EM applications, they are used for the design and deployment of conventional radio systems, field prediction for network planning, and localization~\cite{b0}. Recently, dynamic ray tracing has been gaining popularity with the deployment of 5G networks, the use of these technologies for autonomous driving, and the development of wireless systems for safety and traffic control~\cite{b1}. Dynamic ray tracing in EM simulators is used to predict wave propagation in large environments with moving objects like vehicles or device users. A key issue in these applications is that dynamic environments are susceptible to attenuation and blockage by obstacles. In addition, Doppler shifts come into effect when the source of EM radiation and the observer are in relative motion, causing a shift in the frequency of the EM radiation detected by the observer. Other factors to be considered in dynamic ray tracing are the Line-of-Sight (LoS) and Non-Line-of-Sight (NLoS) transitions when the moving objects are blocked by different obstacles along their trajectories.

Many techniques have also been proposed to mitigate the effects of channel variability and maintain reliable communications. One approach is to use advanced antenna technologies, such as beam-forming and multiple-input multiple-output (MIMO) systems~\cite{b3}\cite{b4}, to improve the directional sensitivity and spatial resolution of the wireless system. Another approach is to use machine learning or artificial intelligence algorithms to adapt the system parameters in real-time based on the current channel conditions \cite{b5}. 
Some of the key challenges in dynamic ray tracing in urban environments include:
\begin{enumerate}
    \item  Complex propagation environments: These environments consist of various types of obstacles in urban scenes such as concrete buildings, trees, and moving vehicles. Many simulators tend to use comprehensive and site-specific propagation models, and it is difficult to generalize them to large, heterogeneous environments;  
    \item Lack of ground truth measurements data: Due to the complex and dynamic urban environments, it can be challenging to accurately measure the varying signal strengths at different locations over a period of time while also precisely tracking the positions of dynamic obstacles in a large scene; 
    \item High computational complexity: The complexity of current ray tracing methods increases with the size of the environment as well as the number of obstacles. As a result, most  current simulators are limited in terms of dynamic ray tracing capabilities and do not deal with large or complex urban scenes~\cite{b2}.
    \end{enumerate}

\noindent {\bf Main Results:} In this paper, we present a novel approach to perform dynamic EM ray tracing in large urban scenes with multiple receivers. Our approach is based on spatial and temporal coherence techniques that have been proposed to accelerate ray tracing  for acoustic~\cite{b55} and EM simulations~\cite{b6}, We also  present a new method to model channel coherence and spatial consistency in ray tracing models and combine them with coherence methods for faster computations. We also use path caching, backward tracing, and dynamic bounding volume hierarchies to  perform fast ray tracing simulations in large urban scenes. We show that our approach scales with the size of the scene as well as the number of obstacles and dynamic objects.  
The novel components of our work include:
\begin{itemize}
\item We present a fast ray tracing-based dynamic simulator for EM bands for complex urban environments.  
\item We provide efficient methods to model channel coherence, spatial consistency, and Doppler effects with coherent ray tracing to  improve the accuracy of dynamic object propagation.
\item We evaluate the performance and accuracy of our approach on  large dynamic urban environments of size $2*2 km^2$, where we vary the number of receivers from 1 to 10,000. We observe that the running time scales linearly with the area of the scene and the number of receivers. Furthermore, the accuracy is comparable to the discrete model simulation results computed using $WinProp$, which is applied to different static instances.

\end{itemize}
The rest of the  paper is organized as follows: In Section II, we give an overview of prior work and the Dynamic Coherence-Based EM Ray Tracing Simulator (DCEM) \cite{b6}. We present new techniques that can incorporate  channel coherence, spatial consistency, and Doppler effects in the coherence-based ray tracing method for large scenes in Section III. In Section IV, we describe the urban scenes used to evaluate the performance of our approach and how we vary their size and dynamic complexity. We also compare and analyze the simulation results and address performance issues that govern its scalability and accuracy.

\section{ Prior Work }
There is considerable work on developing specialized or site-specific propagation models for urban scenes~\cite{b13,b14,b15}. However, these methods do not generalize to new or arbitrary scenarios~\cite{b16}. Many other widely used algorithms and software systems have also been designed for EM ray tracing.  These include $WinProp$~\cite{b17}, which performs standard ray tracing, intelligent ray tracing, and dominant path models at frequencies up to 100 GHz. This system also supports simulating spatial variability. We generate various static instances or configurations of the scene from our benchmarks, corresponding to particular time instances, and compare the accuracy for these configurations with $WinProp$.
$Wireless Insite$ uses ray tracing methods  coupled with empirical and deterministic models and can handle frequencies up to 100 GHz~\cite{b18}. Many techniques have been proposed to accelerate the computations using dimension reduction and exploiting the parallelism of current CPUs and GPUs. $Cloud RT$ is a cloud-based system that supports frequencies up to $325$ GHz while also supporting different propagation models and scattering objects~\cite{b20}. It uses spatial partitioning hierarchies to accelerate the ray intersections with the objects in the scene. $GEMV^2$ is used to analyze vehicle-to-vehicle channels in large urban environments~\cite{b22} and can also handle tens of thousands of vehicles. It is a MATLAB-based system and is evaluated for frequencies up to $5.9$ GHz. It is not clear whether these systems can automatically handle large dynamic urban scenes with thousands of receivers and high frequencies.

\subsection{Coherence-based Ray Tracing}
Our approach exploits a coherence-based framework and extends DCEM~\cite{b6} to handle large urban models and perform accurate computations in dynamic scenes.
The underlying ray tracing algorithm models direct rays, reflected rays, diffracted rays, and scattered rays  to perform EM propagation simulations. It uses three main characteristics to accelerate the performance in vehicular environments:  (1) backward ray tracing from the receiver in the visibility determination step; (2) the use of bounding volume hierarchy (BVH), a tree structure computed on a set of geometric objects, where all geometric objects that form the leaf nodes of the tree are wrapped in bounding volumes (it updates these dynamic BVHs to reduce the rebuilding cost between successive frames to accelerate the computations); and (3) propagation path caching to exploit spatial and temporal coherence. This path caching reduces the overhead of performing a high number of ray intersections with the objects in the scene. In practice, the use of coherence reduces the computation time by $36\%$ compared to $GEMV^2$ and by $30\%$ compared to WinProp while maintaining similar prediction accuracy\cite{b6}. 

\subsection{Improvements Based on Previous DCEM}
In this work, we extend the previous engine \cite{b6} to large urban scenes with multiple receivers and address its scalability. We introduce spatial consistency and channel correlation to improve ray tracing simulation's efficiency and accuracy in dynamic scenarios and consider the Doppler effects under such settings. More details are discussed in the next section.
\section{Dynamic EM Ray Tracing for Complex Scenes}
In this section, we present novel efficient methods to model the spatial consistency, channel correlation, and Doppler effect with the coherence-based ray tracing approach. It is important to model these effects in large, dynamic urban scenes for accurate EM propagation.

\subsection{Spatial consistency}
\begin{figure}[htbp]
\centerline{\includegraphics[scale=1.5]{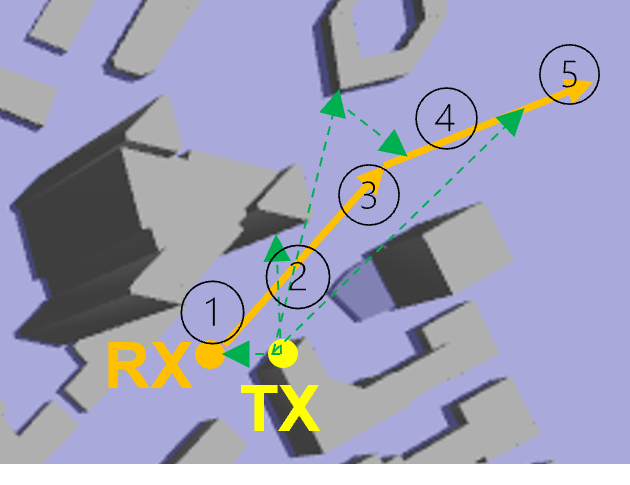}}
\caption{A trajectory used for simulating propagation is divided into five segments based on maintaining spatial consistency. The transmitter is shown in yellow and is at a fixed location. In this case, the moving receiver is along the orange route. The green dashed lines correspond to LoS paths and primary reflection paths, which are used to perform segmentation.}
\label{fig4}
\end{figure}

Spatial consistency~\cite{b7} refers to the degree to which the channel characteristics remain consistent over a certain spatial distance, typically in a local area (e.g., within 10-15m)~\cite{b8}. By analyzing the channel characteristics along the moving object's trajectory, it is possible to determine the degree of spatial consistency as well as how well the channel characteristics evolve along the trajectory. In our coherence-based framework, we use large-scale parameters such as distance and LoS/NLoS conditions, as well as small-scale parameters such as the angle of arrivals, power, phase, and delay of the propagation paths to determine the spatial consistency. Based on these computations, we further divide each object's trajectory into several segments. The channels are considered highly correlated within each segment and the same channel model can be used. Furthermore, the BVH update within each segmentation can be performed efficiently since the changes in BVH computations are relatively small. In most cases, only the subtrees need to be rebuilt. Given the updated hierarchy, spatial consistency would be computed and updated based on the approach described in \cite{b9}. 
An illustration of trajectory segmentation is shown in Fig.~\ref{fig4}. 
\subsection{Channel Correlation Function and Coherence Time}
The channel correlation function is a mathematical function that describes the statistical correlation between two signals that are transmitted over the wireless channel at different times. Specifically, it describes how the channel affects the correlation between the signals and how this correlation changes over time. This is important for large dynamic environments with moving receivers. The correlation function can be defined as:
\begin{equation}
   R_{|h|}(\tau)=\frac{\mathop{\mathbb{E}[g(t)g(t+\tau)]}-\mathbb{E}[g(t)]^2}{\mathbb{E}[g(t)^2]\mathbb{E}[g(t)]^2}
\end{equation}
where $g(t)=|h(t)|$, $h(t)$ is the channel coefficient and $\mathbb{E}$ denotes the expectation operator. With the correlation function defined, a simplified equation for computing coherence time is:
 \begin{equation}
     T_{c}(\theta)=\frac{\sqrt{1/R^4-1}}{2\pi f_D\theta^2}
 \end{equation}
 where $R$ is a predefined threshold value of channel correlation, $f_D$ is the maximum Doppler shift frequency, and $\theta^2$ can be approximated by the angle of the object's moving direction and propagation path between transmitter and receiver.

\begin{figure}[htbp]
\centerline{\includegraphics[scale=1.1]{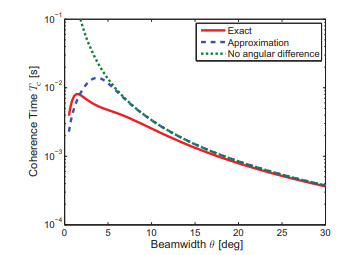}}
\caption{Channel coherence time approximation when the angle between moving direction and propagation path is small. “Approximation” and “No angular difference” are different models proposed in \cite{b10}. We can combine this approximation with path caching in our coherence-based formulation.}
\label{fig5}
\end{figure}

 The coherence time is the length of time over which the channel correlation function remains significant. In other words, it is the time scale over which the channel characteristics remain relatively constant. The coherence time can determine the length of time over which the system can accurately predict the behavior of the channel. The detailed derivations can be found in \cite{b10}. An evaluation plot of the approximated computation of coherence time is shown in Fig.~\ref{fig5}. 
 
 We compute coherence time within each segment in  the channel and update it accordingly. During the time duration corresponding to this coherence time, we assume that the positions of the receivers do not change significantly. As a result, we can use path caching  to accelerate ray paths and intersection computations, i.e., valid paths are stored in the cache and updated until they are removed. This deletion of a path from the visibility hash table is performed to lower the number of visibility rays cast within the coherence time.

\subsection{Doppler Effects}
The Doppler effect is a well-known phenomenon in physics that describes the change in frequency of a wave due to the relative motion between the wave source and the observer. In particular, the Doppler effect in EM propagation or mmWave systems is caused by the relative motion between the transmitter and the receiver due to the high mobility of the devices or vehicles, which results in frequency shifts. It is important to model these effects in dynamic environments for accurate EM propagation. In multi-path propagation, two signals from the same transmitter, coming from two paths and reflected by moving objects with different speeds, might arrive at the receiver as frequency-shifted signals interfere and create fading, similar to phase-shifted signals that cancel each other out by destructive interference. To demonstrate the capacity of dynamic simulations, it is critical to include the mean Doppler shifts and Doppler spread in system performance evaluation \cite{b23}. The Doppler shift of the LoS can be calculated by:
 \begin{equation}
     f_d(t)=f_c\frac{v}{c}cos\theta
 \end{equation}
where $f_c$ is the central frequency, $v$ is the object's moving speed, $c$ is the light speed, and $\theta$ is the angle between the object's moving direction and propagation path.

In our formulation, we assume that the base station is static. As a result, we track the moving speeds of objects along their trajectories and compute the Doppler shifts and Doppler spread to adapt proper propagation models to generate received powers and delays. The results are shown in Section V.

\section{Simulation and Evaluation}
We perform our simulation in complex urban environments corresponding to a city with multiple blocks, many buildings, and moving vehicles. We model a base station (BS) covering a certain area for signal broadcasting in an urban environment with many moving objects including high-speed/low-speed vehicles and pedestrians. Our goal is to evaluate the signal coverage simulation in dense situations when a lot of receivers are in close proximity. This may happen during traffic. We perform simulations to address the scalability and performance of our system in terms of the number, speed, and distribution of the objects. 

\begin{figure}[htbp]
\centerline{\includegraphics[scale=0.45]{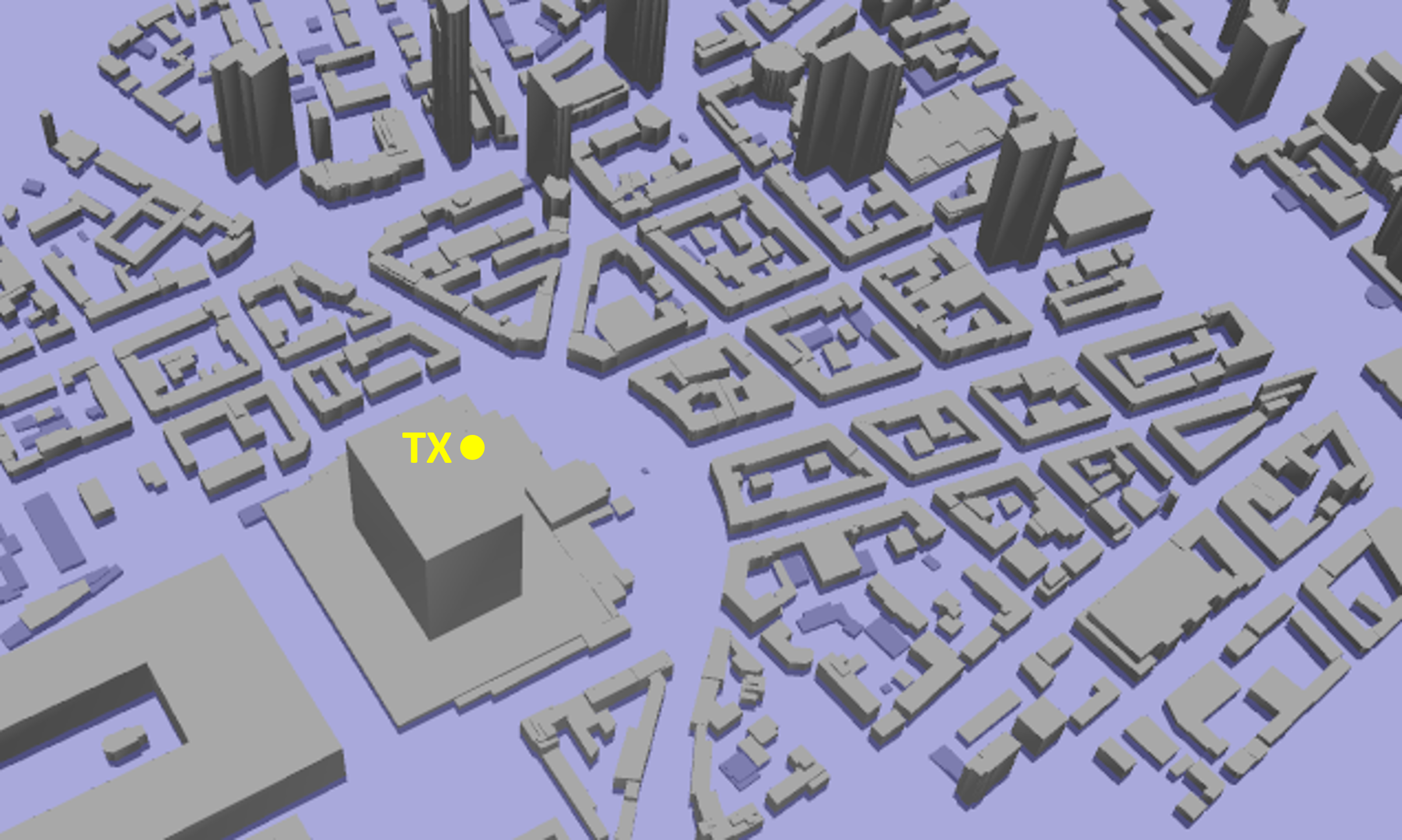}}
\caption{A screenshot of the 3D city model used to evaluate our performance. It is a complex urban scene with many buildings and vehicles. The yellow dot is the fixed location of TX (source). We show that our method can perform dynamic ray tracing simulations in this environment with up to 10,000 moving obstacles, with high accuracy and fast computation time.}
\label{fig3}
\end{figure}

\begin{figure}[htbp]
\centerline{\includegraphics[scale=0.33]{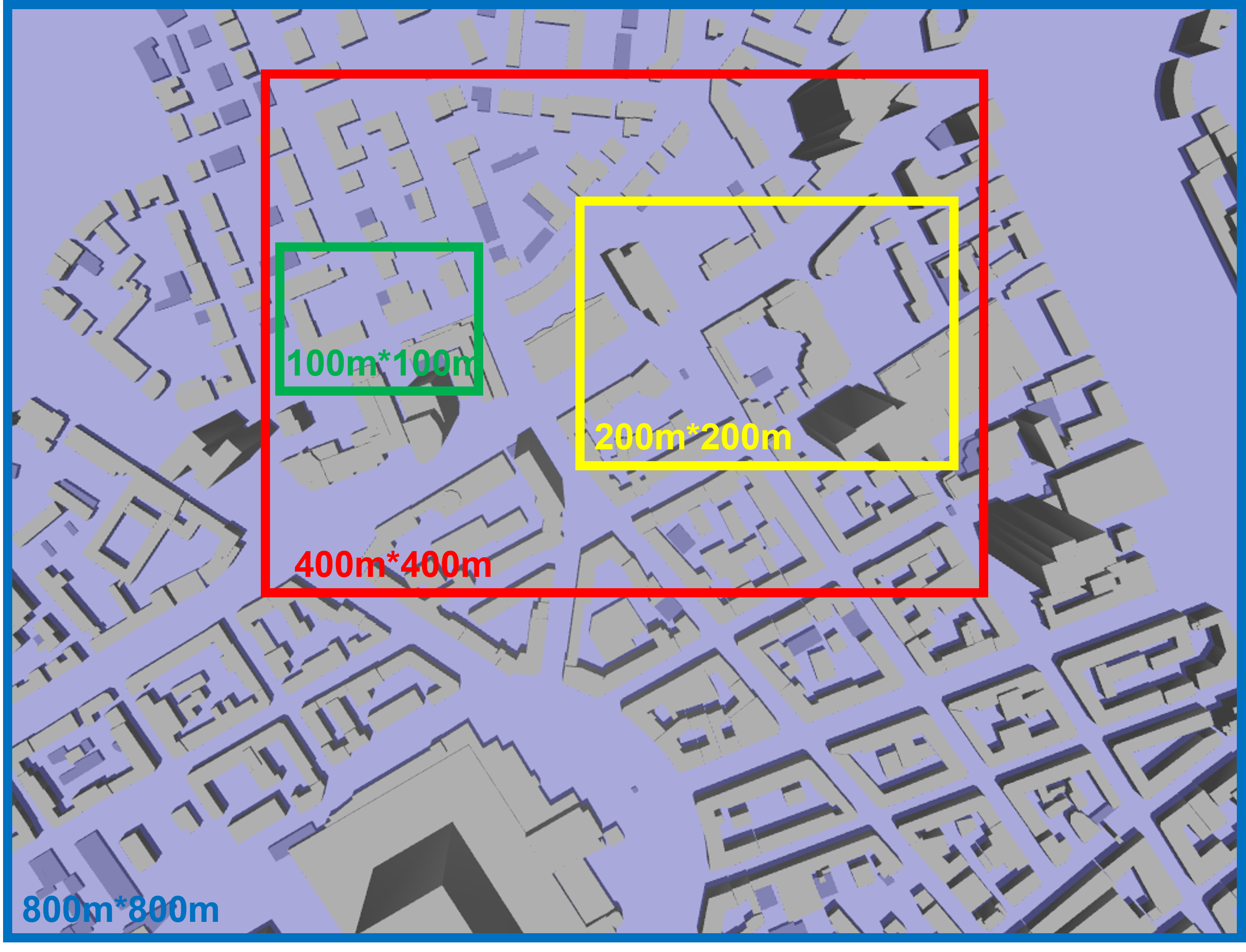}}
\caption{Illustration of various simulation environment sizes. We perform ray tracing simulations in different areas of varying sizes in the modeled environments with a fixed TX location and moving RX with pre-generated trajectories inside each simulation area. The boxes in different colors indicate areas of simulation with different sizes. We run simulations in 10 randomly selected areas of the same size and record the computation time to compute the average performance.}
\label{fig7}
\end{figure}

The simulated area is a $2*2 km^2$ urban city environment with 750 modeled building blocks. A top view of the modeled city or the urban environment is shown in Fig.~\ref{fig3}. The transmitter (TX) is fixed on the rooftop (100m) of a high-rise building, as shown in Fig.~\ref{fig3}. The receivers correspond to the vehicles and moving objects in the urban environment with predefined trajectories. However, our ray tracing-based simulator does not make any assumptions about these trajectories. We evaluate the performance of our simulation algorithm by changing the size of the urban environment, as shown with boxes of different colors highlighted in Fig.~\ref{fig7}. We consider areas corresponding to $10K$ $meter^2$, $40K$ $meter"2$, and $160K$ $meter^2$. Basically, each dimension of the square region increases linearly.

\begin{figure}[htbp]
\centerline{\includegraphics[scale=0.3]{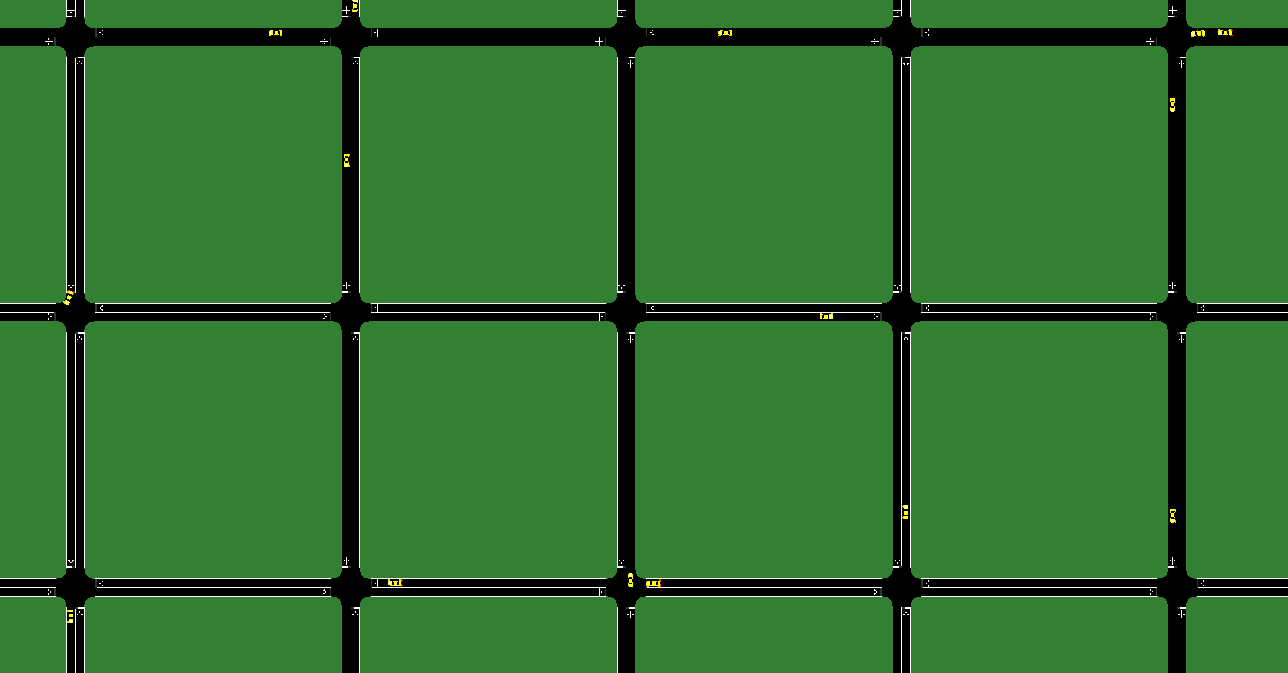}}
\caption{SUMO visualization in the urban environment: We export the traffic route information from SUMO and feed it into a dynamic ray tracing simulator. The moving objects, corresponding to yellow cars, are generated in a random and automatic manner. We use the positions of these receivers to compute the signal strength at the corresponding location.}
\label{fig6}
\end{figure}

We use SUMO (Simulation of Urban Mobility)\footnote{https://www.eclipse.org/sumo/}, which is a traffic simulation software designed for modeling and analyzing traffic flow in urban areas. We use SUMO to model and generate the trajectories of moving objects that correspond to receivers in urban environments. The trajectory can be generated automatically by SUMO's built-in routing algorithms. Once the trajectory is generated, it can be used to simulate the movement of the vehicles considering the dynamic flow of the traffic and the interactions with other vehicles in the environment.  Fig.~\ref{fig6} shows the traffic generated by SUMO in a sample network. The black lines indicate streets, the green regions correspond to non-traffic areas, and the small yellow cars refer to moving traffic in the scene.

The following table summarizes the material properties of the objects in the urban used environment to generate the propagation results.  The transmitter is set to broadcast signals at frequency $3GHz$ and at power $50dBm$. The antenna is assumed to be ideal and omnidirectional. Since the buildings are rather thick in general, we do not consider penetration or transmission through the building walls.
\begin{table}[htbp]
\caption{Material properties used in outdoor scenes}
\resizebox{\columnwidth}{!}{\begin{tabular}{|l|l|l|l|l|}
\hline
\textbf{Material} & \textbf{\textit{Thickness(mm)}}& \textbf{\textit{Reflection coefficient}}& \textbf{\textit{Penetration loss(dB)}} \\
\hline
Wall  & /& 0.8& / \\
\hline
Metal  & 10& 0.9& 10 \\
\hline
\end{tabular}
\label{tab2}
}
\end{table}

\section{Results and Analysis}
In this section, we present the running time performance of our EM simulator in complex urban scenes and compare its accuracy with the prior method. 
\subsection{Scalability}
\begin{figure}[htbp]
\centerline{\includegraphics[scale=0.75]{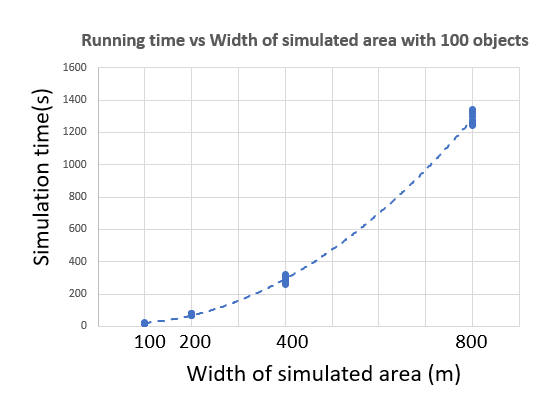}}
\caption{The plot of simulation time vs. width of simulated area with 100 objects. The X-axis represents the width of simulated areas in meters and the Y-axis is the running time in seconds. The simulation areas of different sizes are randomly selected for the position of the objects. The largest area of size $800*800m^2$ is generated and the smaller sized areas are the subsets of the larger areas. 
We observe that, when the area width grows from $100$m to $800$m, the running time grows from around $20s$ to $1300s$, which is almost a quadratic function of the width and a linear function of the area.}
\label{fig1}
\end{figure}

\begin{figure}[htbp]
\centerline{\includegraphics[scale=0.75]{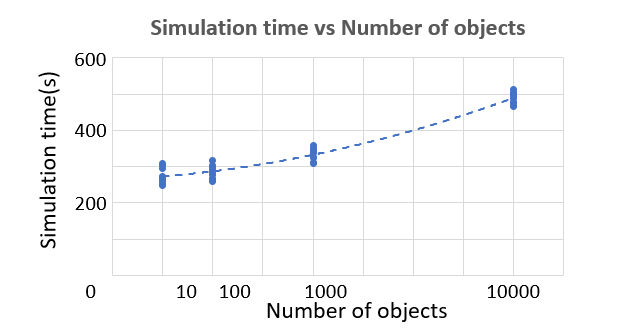}}
\caption{The plot of simulation time vs. the number of objects. The X-axis represents the number of moving objects in a random $400*400m^2$ scene and the Y-axis is the running time in seconds. We chose 10 different regions in the urban environment and compute an average running time over them. It is observed that when the number of moving objects grows from $10$ to $10,000$, the running time increases from an average of $270$ sec to $600$ sec, which is a sub-linear growth. This is due to three factors: (1) the underlying complexity of the modeled 3D environment and our use of a bounding volume hierarchy to accelerate ray tracing computations;  (2) the moving objects being modeled as a standard-size box of metal, and our use of a fixed representation for each such object.; (3) the use of acceleration methods such as spatial consistency, coherence time, and coherence-based ray tracing, which significantly reduce the computation complexity.}
\label{fig2}
\end{figure}

We evaluate the performance of the simulator as a function of the area of the urban scene and the number of receivers.
We highlight the running time comparison plots for scenarios corresponding to:
\begin{itemize}
\item Given a fixed area of the urban scene, we increase the number of objects linearly. The performance is shown in Fig.~\ref{fig2}. We observe that the running time increases at a sub-linear rate with respect to the number of objects in the scene.
\item We increase the area of the urban environment while using the same number of objects or moving obstacles in the scene. The total running time as a function of the width of the simulated area is shown in Fig.~\ref{fig1}. It increases as a non-linear function of the width but is a sub-linear function of the area, which can be computed as $Area=Width^2$ for these square regions.
\end{itemize}

\subsection{Accuracy}
\begin{figure}[htbp]
\centerline{\includegraphics[scale=0.55]{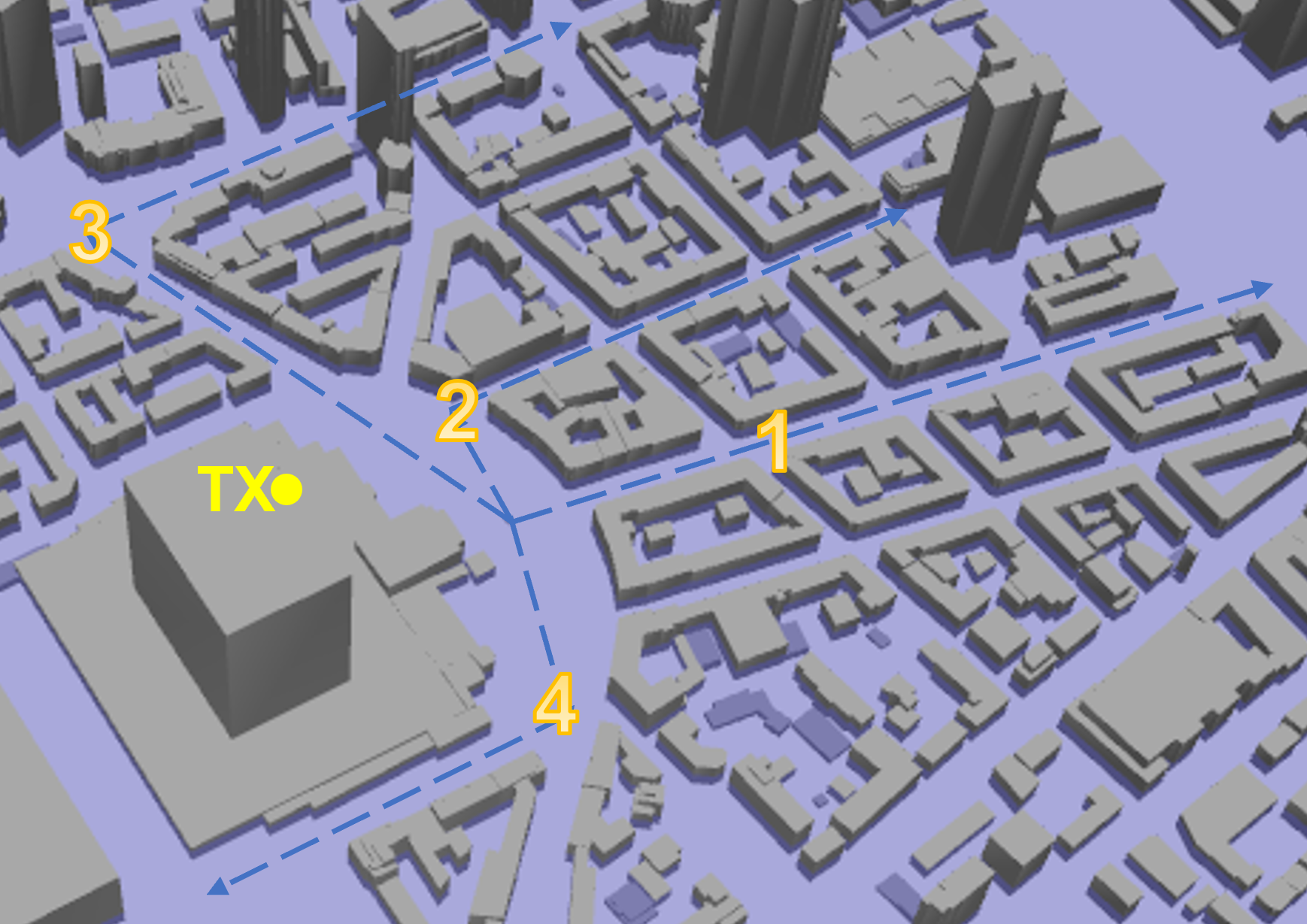}}
\caption{ We highlight the trajectories simulated using our dynamic ray tracing algorithm as well as Winprop in complex urban scenes. Our approach considers various trajectories. In particular, 
the blue dashed lines with indices in orange show the moving trajectories. In Trajectory 1, it is mostly due to LoS computation; in Trajectory 2, the LoS and NLoS computations are mixed after the turning point; in Trajectory 3, it is most due to NLoS computations after entering the area of buildings on the right; in Trajectory 4, the moving object is close to the TX, but is blocked by the building.}
\label{fig8}
\end{figure}

\begin{figure}[htbp]
\centerline{\includegraphics[scale=0.26]{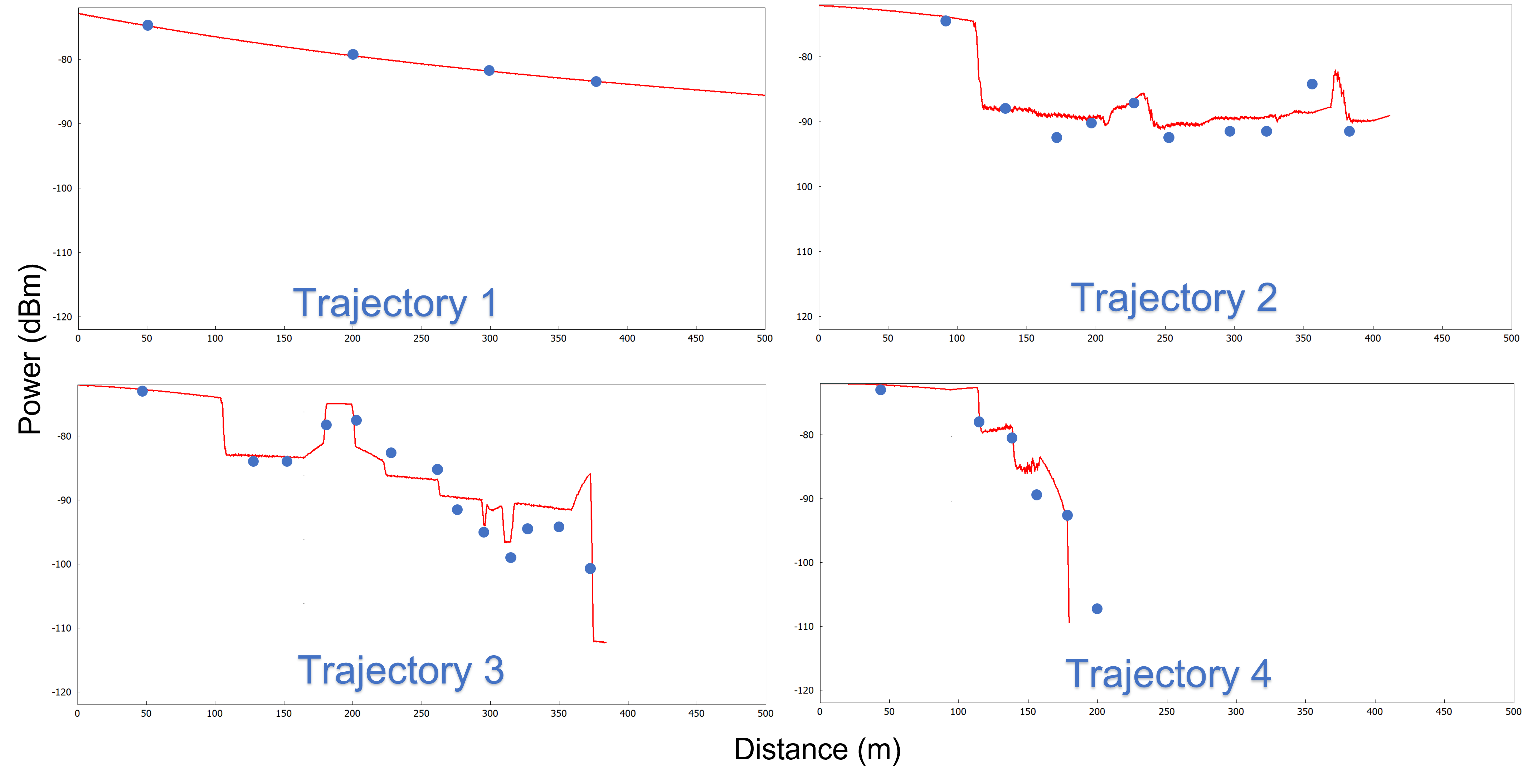}}
\caption{{\bf Accuracy Comparisons with WinProp:} These plots highlight the  received power using our simulator along  four routes highlighted in Fig.~\ref{fig8}. The plots generated using our dynamic simulator are shown in red,  while the static simulations generated using a given configuration of the obstacles and receivers are generated using  WinProp are shown with blue dots. This comparison shows that the accuracy of our dynamic simulator aligns well with those computed using WinProp for different trajectories. We see a very  high match in Trajectory 1, which corresponds to most LOS computations. We see small differences as we model higher-order reflections and Doppler effects in our formulation. 
}
\label{fig9}
\end{figure}

\begin{figure}[htbp]
\centerline{\includegraphics[scale=0.42]{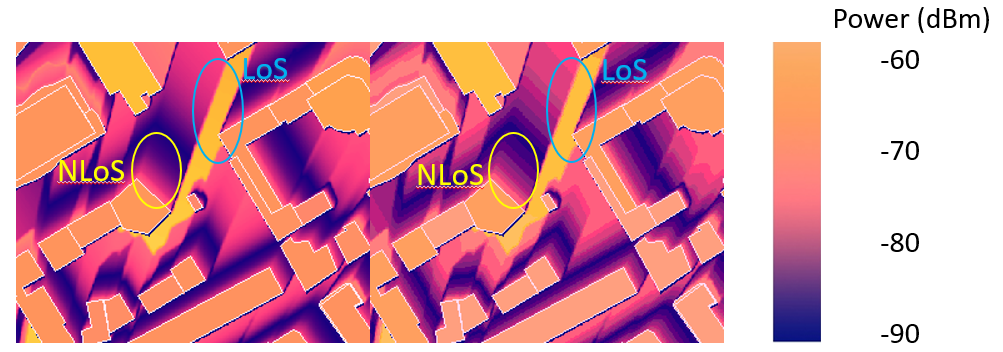}}
\caption{ Static simulation (left) and dynamic results (right). It is observed that the LoS channels (indicated in blue circles) are mostly the same in both simulations, while more variations can be seen around obstacles in dynamic simulations (indicated in yellow circles). This is because when the transmitter (at the same location as in Fig 8, not included in this small area) illuminates the buildings/obstacles from the top, the shadow area caused by building blockages could be characterized into several states, where we apply different propagation models to each state, leading to a clear boundary between different states. }
\label{fig10}
\end{figure}

\begin{figure}[htbp]
\centerline{\includegraphics[scale=0.36]{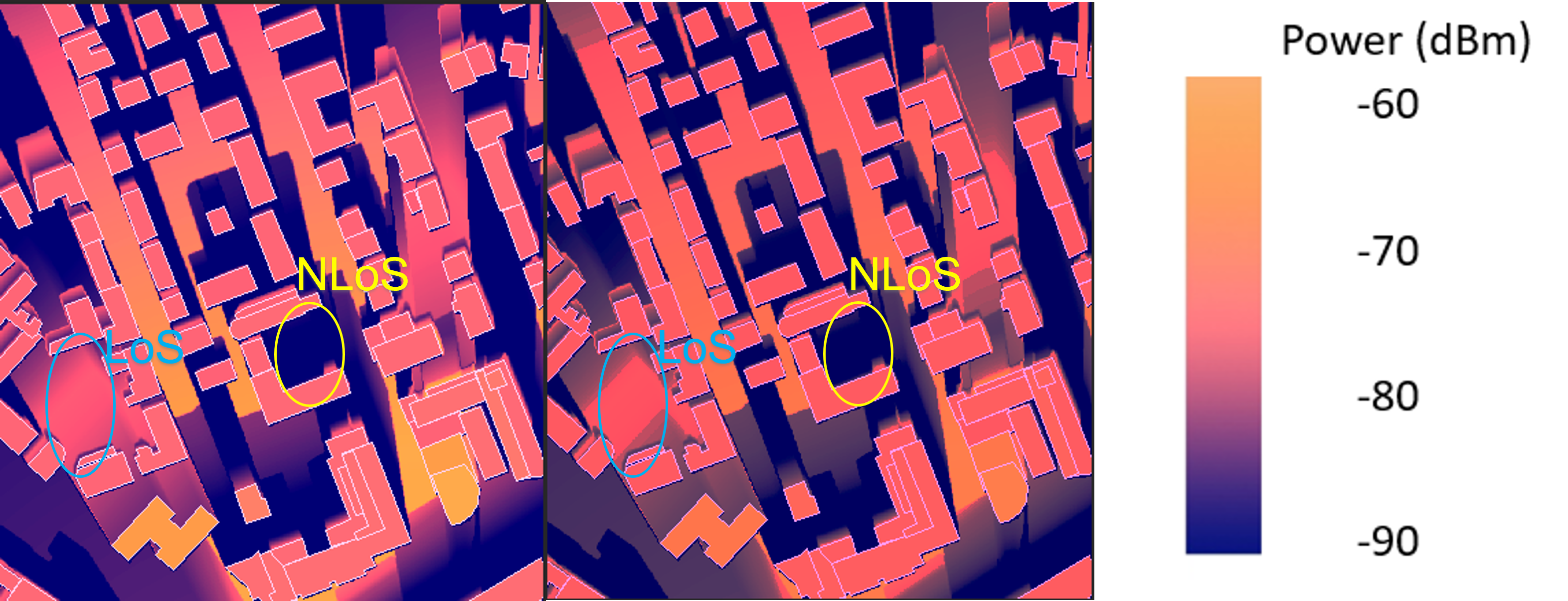}}
\caption{ Static simulation (left) and dynamic results (right). It is observed that the LoS channels (indicated in blue circles) and the NLoS channels (indicated in yellow circles) are mostly the same in both simulations. In this comparison, more NLoS predictions (low powers in dark blue ) are presented. }
\label{fig11}
\end{figure}

There is no available ground truth data for EM propagation in complex dynamic environments. Furthermore, many software systems either do not support large dynamic scenes or handle high frequencies. As a result, there is no prior work that can be used to compare the accuracy or validate the results from our simulation. Even commercial systems like WinProp do not support dynamic simulations in outdoor environments. Instead, we evaluate the accuracy of our simulator  by computing the predicted power at different locations of a given moving object and compare the results with WinProp predictions under static conditions or configurations at these locations. We compare their accuracy for different trajectories highlighted in Fig.~\ref{fig8}. We show the power plots computed using our dynamic ray tracer along the trajectories (shown in red) with the static configurations (shown in blue) using WinProp in Fig.~\ref{fig9}. As indicated in Fig.~\ref{fig9}, the dynamic predictions align well with the discrete static predictions. The average absolute power difference of the four trajectories is 2.3 dB. For Trajectory 1, it is a simple LoS channel and thus fewer simulated comparisons are performed; the results match very well. We see small differences in other trajectory simulations, which come from the Doppler effects and higher-order reflections.
A heatmap comparison between static and dynamic simulations at a certain location is shown in  Fig.~\ref{fig10}. It is observed that for most areas, the simulations align well in terms of signal propagation effects and simulated powers, except for some blocked areas, which might result from different diffraction models and higher-order reflections.
Another heatmap comparison in Fig.~\ref{fig11} addresses the possibility of accurate simulations in most NLoS simulations. One more series of heatmap evaluations on a moving path is shown in the Appendix.

\subsection{Running Time in Static and Dynamic Setups}
We compare the static and dynamic algorithm running times along Trajectory 2. The object is set to $10m/s$ along the $400m$ route. The static algorithm runs 40 times to cover the route while the dynamic algorithm runs just 1 time. The running time improvement is musch larger in 1 dynamic run than in multiple static simulation runs (shown in Table II).
\begin{table}[htbp]
\caption{Running time comparison}
\resizebox{\columnwidth}{!}{\begin{tabular}{|l|l|l|l|l|}
\hline
\textbf{First Static Run (s)} & \textbf{Second Static Run (s)}& \textbf{First Dynamic Run (s)}& \textbf{Second Dynamic Run (s)} \\
\hline
4726  & 4645 & 244& 252 \\
\hline
\end{tabular}
\label{tab1}
}
\end{table}
\subsection{The Mean Doppler Shift and Doppler Spread}
The following table shows the mean value of Doppler shift and RMS Doppler spread at different moving speeds of the objects at 30GHz bands. It can be observed that the Doppler shift is low when the LoS path is perpendicular to the moving direction and high when the distance is large. The RMS Doppler spread is associated with the angular spread. The results are shown in Table III.
\begin{table}[htbp]
\caption{The mean Doppler shift and RMS Doppler spread}
\resizebox{\columnwidth}{!}{\begin{tabular}{|l|l|l|l|l|}
\hline
\textbf{\textit{Moving speed (m/s)}} & \textbf{\textit{Distance (m)}}& \textbf{\textit{Mean of Doppler shift (Hz)}}& \textbf{\textit{RMS Doppler spread (Hz)}} \\
\hline
10  & 10& 5.4&  61.0\\
\hline
10  & 300& 376.0& 0.07 \\
\hline
50  & 10& 27.0& 305.0 \\
\hline
50  & 300& 1880.0& 0.34 \\
\hline
\end{tabular}
\label{tab3}
}
\end{table}

\section{Conclusion and Limitations}
We present a fast method for EM ray tracing in dynamic scenes with many receivers. Our approach uses a coherence-based ray tracer and models spatial consistency, channel correlation, and Doppler effects for large dynamic scenes. We have evaluated the performance on large urban scenarios and show that its running time increases linearly with the area of the scene and the number of obstacles. We also compare the accuracy with WinProp. 

Our approach has some limitations. It is hard to perform accuracy analysis or validation due to lack of ground truth data. It is non-trivial to accurately predict many dynamic effects such as Doppler Effect and spatial consistency in complex environments with dynamic objects.  We only simulated concrete and metal materials in this work, so we need to consider the variations in the material properties when measurement data are available. We also need to further evaluate our approach in other environments and include more effects such as foliage and vegetation.

\section{Appendix}
\begin{figure}[htbp]
\centerline{\includegraphics[scale=0.35]{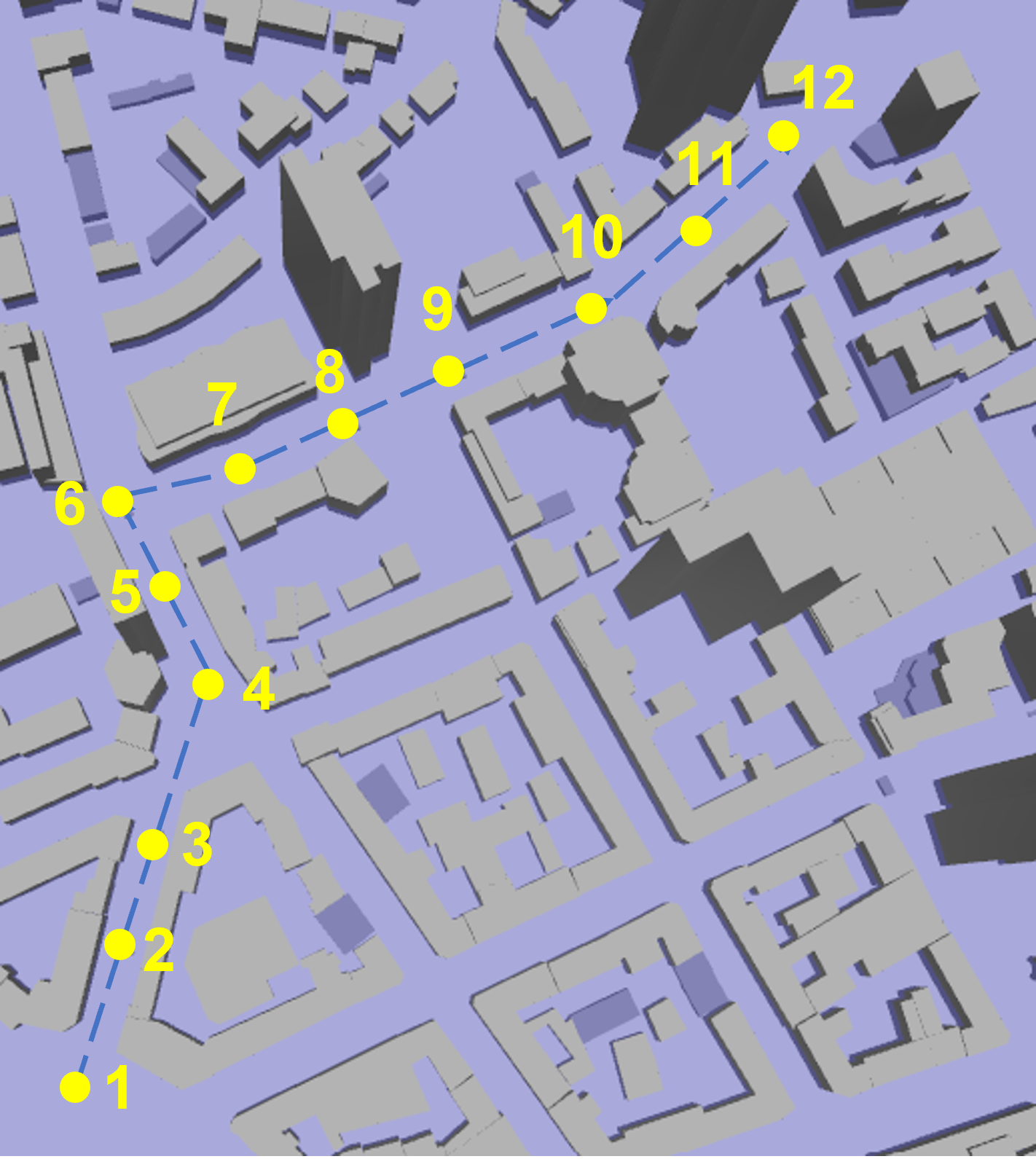}}
\caption{ The simulation path is shown in a blue dashed line and the selected positions are indicated in yellow dots.}
\label{fig12}
\end{figure}

\begin{figure}[htbp]
\centerline{\includegraphics[scale=0.4]{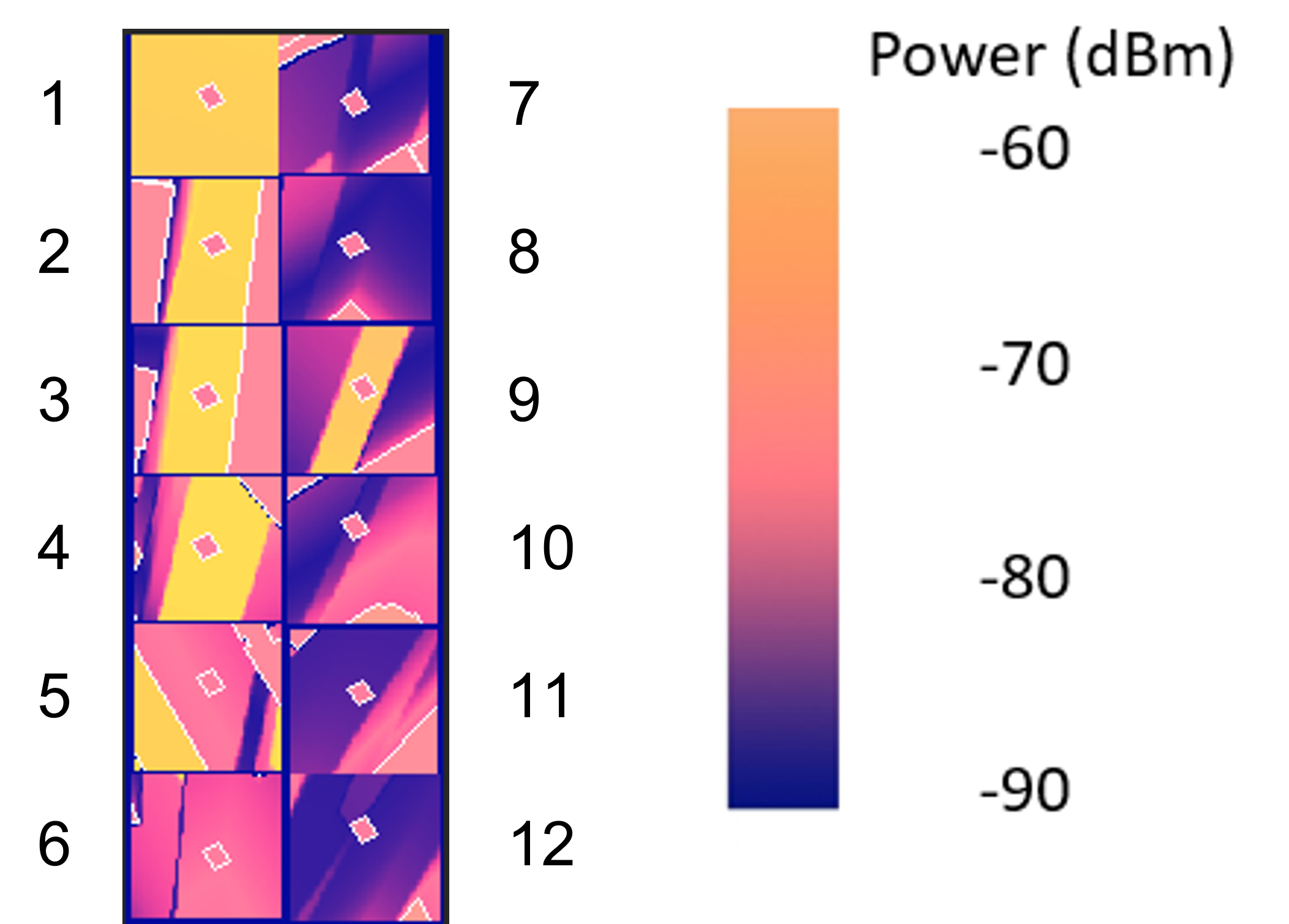}}
\caption{ Static results from WinProp: We observe the surroundings and power changes at different locations.}
\label{fig13}
\end{figure}

\begin{figure}[htbp]
\centerline{\includegraphics[scale=0.4]{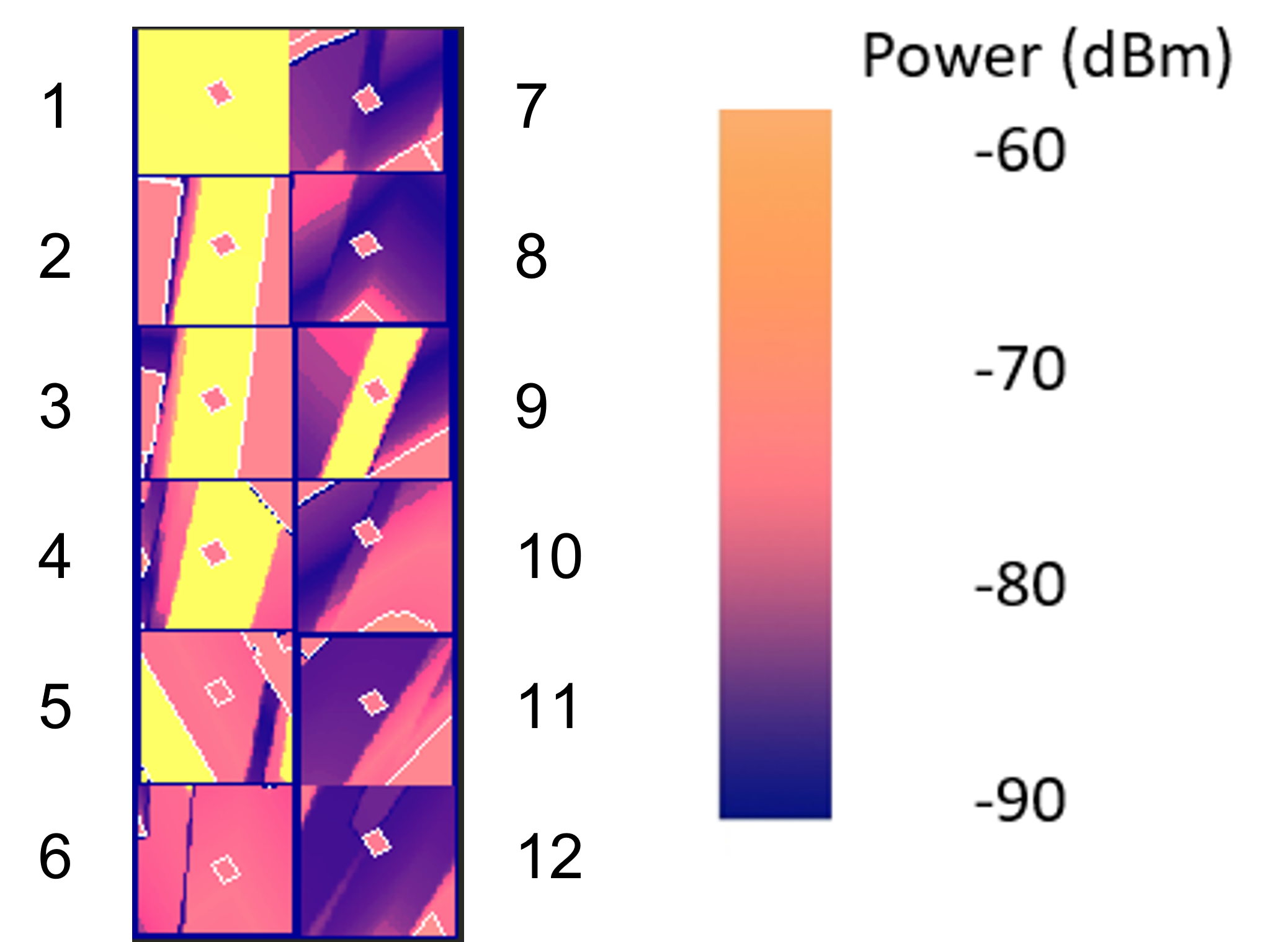}}
\caption{ Dynamic results from our work: We see predictions close to those in Fig.~\ref{fig13}. Some differences are observed in the heatmaps of positions after 8. Considering their locations, these might be the shadow areas created by building blockages, where different propagation models are applied to each state, leading to clearer boundaries.}
\label{fig14}
\end{figure}

We show a series of heatmap comparisons along a moving path between static and dynamic simulations from WinProp and DCEM here. Fig.~\ref{fig12} illustrates the path and the selected positions for static and dynamic simulation comparisons. Fig.~\ref{fig13} and Fig.~\ref{fig14} provide the static and dynamic simulation heatmaps, respectively. The indices beside each small heatmap in Fig.~\ref{fig13} and Fig.~\ref{fig14} correspond to the numbered positions in Fig.~\ref{fig12}. The channel conditions change from LoS (1-4) to mixed (5-9) and mostly NLoS (10-12). The results show that the dynamic simulations align well with static ones at most selected positions.

\begin{thebibliography}{00}
\bibitem{b0} Yun, Zhengqing, and Magdy F. Iskander. "Ray tracing for radio propagation modeling: Principles and applications." IEEE access 3 (2015): 1089-1100.
\bibitem{b1} Bilibashi, D., E. M. Vitucci, and V. Degli-Esposti. "Dynamic ray tracing: Introduction and concept." 2020 14th European Conference on Antennas and Propagation (EuCAP). IEEE, 2020.
\bibitem{b2} Bilibashi, Denis, Enrico M. Vitucci, and Vittorio Degli-Esposti. "On Dynamic Ray Tracing and Anticipative Channel Prediction for Dynamic Environments." arXiv preprint arXiv:2205.00237 (2022).
\bibitem{b3} He, Jiguang, et al. "Adaptive beamforming design for mmWave RIS-aided joint localization and communication." 2020 IEEE Wireless Communications and Networking Conference Workshops (WCNCW). IEEE, 2020.
\bibitem{b4} Wiffen, Fred, et al. "Comparison of OTFS and OFDM in ray launched sub-6 GHz and mmWave line-of-sight mobility channels." 2018 IEEE 29th Annual International Symposium on Personal, Indoor and Mobile Radio Communications (PIMRC). IEEE, 2018.
\bibitem{b5} Alkhateeb, Ahmed. "DeepMIMO: A generic deep learning dataset for millimeter wave and massive MIMO applications." arXiv preprint arXiv:1902.06435 (2019).
\bibitem{b55} Schissler, Carl, Ravish Mehra, and Dinesh Manocha. "High-order diffraction and diffuse reflections for interactive sound propagation in large environments." ACM Transactions on Graphics (TOG) 33, no. 4 (2014): 1-12
\bibitem{b6} Wang, Ruichen, and Dinesh Manocha. "Dynamic Coherence-Based EM Ray Tracing Simulations in Vehicular Environments." 2022 IEEE 95th Vehicular Technology Conference:(VTC2022-Spring). IEEE, 2022.

\bibitem{b7} Ju, Shihao, and Theodore S. Rappaport. "Simulating motion-incorporating spatial consistency into NYUSIM channel model." 2018 IEEE 88th vehicular technology conference (VTC-Fall). IEEE, 2018.
\bibitem{b8} Ju, Shihao, and Theodore S. Rappaport. "Millimeter-wave extended NYUSIM channel model for spatial consistency." 2018 IEEE Global Communications Conference (GLOBECOM). IEEE, 2018.
\bibitem{b9} Ju, Shihao, et al. "A millimeter-wave channel simulator NYUSIM with spatial consistency and human blockage." 2019 IEEE global communications conference (GLOBECOM). IEEE, 2019.

\bibitem{b10} Va, Vutha, and Robert W. Heath. "Basic relationship between channel coherence time and beamwidth in vehicular channels." 2015 IEEE 82nd Vehicular Technology Conference (VTC2015-Fall). IEEE, 2015.
\bibitem{b13} Rappaport, Theodore S., et al. ``Millimeter wave mobile communications for 5G cellular: It will work!." IEEE access 1 (2013): 335-349.
\bibitem{b14} Degli-Esposti, Vittorio, et al. ``Measurement and modelling of scattering from buildings." IEEE Transactions on Antennas and Propagation 55.1 (2007): 143-153.
\bibitem{b15} MacCartney Jr, George R., et al. ``Millimeter wave wireless communications: New results for rural connectivity." Proceedings of the 5th workshop on all things cellular: operations, applications and challenges. 2016.
\bibitem{b16} Nguyen, Huan Cong, et al. ``Evaluation of empirical ray-tracing model for an urban outdoor scenario at 73 GHz E-band." 2014 IEEE 80th Vehicular Technology Conference (VTC2014-Fall). IEEE, 2014.
\bibitem{b17} ``WinProp: Wireless connectivity". Accessed on Apr 8, 2022. [Online]. https://www.altair.com/electromagnetics-applications/
\bibitem{b18} ``Wireless Insite 3D Wireless Prediction Software". Accessed on Apr 8, 2022. [Online]. https://www.remcom.com/wireless-insite-em-propagation-software/
\bibitem{b20} ``CLoudRT: High Performance Antenna, Propagation and Channel modeling Platform". Accessed on Apr 8, 2022. [Online].  http://cn.raytracer.cloud/
\bibitem{b22} ``$GEMV^2$: geometry-based, efficient propagation model for vehicle-to-vehicle (V2V) and vehicle-to-infrastructure (V2I) communication". Accessed on Apr 8, 2022. [Online]. http://vehicle2x.net/
\bibitem{b23} Tang, Pei. "Channel characteristics for 5G systems in urban rail viaduct based on ray-tracing." 2021 4th International Seminar on Research of Information Technology and Intelligent Systems (ISRITI). IEEE, 2021.
\end{thebibliography}
\end{document}